\documentclass[prb,twocolumn,nobalancelastpage]{revtex4}
\usepackage{epsfig,amsmath}

\begin{document}

\draft
\title{Superconducting properties of nanocrystalline MgB$_2$ thin films
made by an {\em in situ} annealing process}

\author{X. H. Zeng}
\author{A. Sukiasyan}
\author{X. X. Xi}
\author{Y. F. Hu}
\author{E. Wertz}
\author{Qi Li}

\address{Department of Physics, The Pennsylvania State University,
University Park, PA 16802}

\author{W. Tian}
\author{H. P. Sun}
\author{X. Q. Pan}
\address{Department of Materials Science and
Engineering, The University of Michigan, Ann Arbor, MI 48109}

\author{J. Lettieri}
\author{D. G. Schlom}
\author{C. O. Brubaker}
\author{Zi-Kui Liu}

\address{Department of Materials Science and Engineering,
The Pennsylvania State University, University Park, PA 16802}

\author{Qiang Li}
\address{Division of Materials and Chemical Science, Brookhaven National Laboratory,
Upton, NY 11973}

\author{ }
\address{ }

\begin {abstract}
We have studied the structural and superconducting properties of
MgB$_2$ thin films made by pulsed laser deposition followed by
{\em in situ} annealing. The cross-sectional transmission
electron microscopy reveals a nanocrystalline mixture of textured
MgO and MgB$_2$ with very small grain sizes. A zero-resistance
transition temperature ($T_{c0}$) of 34\,K and a zero-field
critical current density ($J_c$) of $1.3 \times 10^6$\,A/cm$^2$
were obtained. The irreversibility field was $\sim$ 8 T at low
temperatures, although severe pinning instability was observed.
These bulk-like superconducting properties show that the {\em in
situ} deposition process can be a viable candidate for MgB$_2$
Josephson junction technologies.
\end{abstract}

\maketitle

Besides its potential for high-current and high-field
applications, the newly discovered superconductor MgB$_2$
\cite{Nagamatsu01,Cava01} has also stimulated great interest in
its applications in microelectronics. It has been shown that
MgB$_2$ is a phonon-mediated BCS superconductor \cite{Budko01}
with an energy gap of 5.2\,meV at 4.2 K \cite{Karapetrov01} and a
coherence length of 50\,\AA. \cite{Finnemore01} Its grain
boundaries do not have large detrimental effects on
superconducting current transport.
\cite{Larbalestier01,Bugoslavsky01} These properties promise that
Josephson junctions of MgB$_2$ may be much easier to fabricate
than those made from the high temperature superconductors. Such
junctions could have the performance of conventional
superconductor junctions, such as Nb and NbN, but operate at a
much higher temperature.

A MgB$_2$ film processing technique compatible with multilayer
depositions is needed for Josephson junction applications.
Currently, two main types of deposition processes have been used.
The first type employs {\em ex situ} annealing of low-temperature
deposited B or Mg+B films at 900 $^{\circ}$C in Mg vapor. The
resultant films exhibit bulk-like $T_{c0} \sim 39$ K
\cite{Shinde01,HYZhai01,WNKang01} and extremely high critical
current density ($\sim$ 10$^7$\,A/cm$^2$ at low
temperatures).\cite{Eom01,Moon01} However, the high-temperature
{\em ex situ} annealing is unlikely to be compatible with
multilayer device fabrications. The second type uses an {\em in
situ} two-step process. Thin films or multilayers of Mg+B or
Mg+MgB$_2$ are deposited at low temperatures, and then annealed
{\em in situ} in the deposition chamber at about 600
$^{\circ}$C.\cite{Blank01,Christen01,Shinde01,Grassano01}
Although this process is potentially more compatible with junction
fabrications, the early reports from various groups on {\em in
situ} MgB$_2$ thin films show lower $T_{c0}$ around or below
25\,K.\cite{Blank01,Christen01,Shinde01,Grassano01} In this paper
we report that high $T_{c0}$ and $J_c$ can be obtained in thin
films made by an {\em in situ} process using pulsed laser
deposition (PLD) from a single target. The structural and
superconducting properties of these films as compared to the {\em
ex situ} annealed films and polycrystalline bulk samples are
discussed.

The MgB$_2$ films were deposited on (0001) Al$_2$O$_3$ substrates
by PLD with an {\it in situ} annealing procedure similar to those
described by Blank {\it et al.}, \cite{Blank01} Christen {\it et
al.}, \cite{Christen01} and Shinde {\it et al.} \cite{Shinde01}.
The PLD targets were prepared by pressing Mg powder with B or
MgB$_2$ powder at room temperature. Some targets were wrapped in
Nb foil and sintered at 600 $^{\circ}$C for 30 min under a mixed
gas of 95\% Ar 5\% H$_2$. Although both targets yielded
comparable films, the results reported in this paper were from an
unsintered target with a Mg:MgB$_2$ molar ratio of 4:1. The films
were deposited at 250 - 300\,$^{\circ}$C in an Ar atmosphere
(99.999\% gas purity) of 120 mTorr. The background vacuum was in
the low to mid 10$^{-7}$ Torr range. The energy density of the
laser beam was 5 J/cm$^2$ and the repetition rate was 5 Hz. The
deposited films were then heated at a rate of
40\,$^{\circ}$C/minute to 630\,$^{\circ}$C and held there for 10
minutes. The atmosphere during the heating and annealing was the
same as during the deposition. After the {\em in situ} annealing,
the films were cooled to room temperature in $\sim$ 20 Torr Ar.
The structure of the films were studied by both x-ray diffraction
and cross-sectional transmission electron microscopy (TEM). TEM
studies were performed in a JEOL 4000 EX microscope operated at
400 kV, providing a point-to-point resolution of 0.17 nm.

In contrast to {\it ex situ} annealed MgB$_2$ films, x-ray
diffraction scans of our {\it in situ} annealed films revealed no
discernable film peaks, indicating that the grain size was
appreciably smaller. This is corroborated by the TEM results shown
in Fig. 1. The dark field image in Fig. 1(a) shows that the film
consists of two layers showing different contrast. Fig. 1(b) is a
selected-area electron diffraction pattern taken from region I
near the film/substrate interface. By measuring the position and
intensity distribution of the diffraction rings, it is determined
that they all belong to the rock salt MgO phase. The MgO
crystallites are very small and textured. Fig. 1(c) is a
diffraction pattern taken from region II close to the film
surface using the same size selected-area aperture as for Fig.
1(b). In addition to the diffraction rings corresponding to the
MgO phase, it also shows diffraction spots corresponding to the
hexagonal MgB$_2$ phase. The diffraction spots indicated by the
circles are due to the (001) planes, while those by the arrow
heads arise from diffraction by the (110) planes of MgB$_2$. The
(021) MgB$_2$ reflections are also detected. The discrete spots
appear in Fig. 1(c) instead of nearly continuous rings in Fig.
1(b), indicating a larger grain size in region II. In both
regions, the result indicates substantial oxygen contamination in
the film. The MgB$_2$ grain size in region I must be less than
about 5 nm to account for the absence of MgB$_2$ rings or spots in
Fig. 1(b).  Additional details on the microstructural analysis
will be published elsewhere.

In Fig. 2(a) we plot the resistivity versus temperature curve for
a 4000\,\AA-thick MgB$_2$ film. It shows a metallic behavior with
a residual resistance ratio, $RRR=R$(300K)/$R$(40K), of 1.4 and
the resistivity at room temperature is $\sim$ 150 $\mu \Omega
\cdot $cm. Compared to high-density bulk samples, where
$RRR=25.3$ and $\rho$(300K)$=9.6 \mu \Omega \cdot
$cm,\cite{Canfield01} the residual resistance ratio of the
MgB$_2$ film is much smaller and the resistivity much higher.
This is likely due to the small grain sizes and existence of MgO
in the film, since precipitates of MgO at the grain boundaries
will act as series-connected resistors to the MgB$_2$ grains. The
superconducting transition temperature of the film, on the other
hand, is close to that of bulk MgB$_2$. The zero resistance
temperatures of the film is 34\,K, which is shown more clearly in
the inset of the figure. The superconducting transition is further
characterized by the ac susceptibility, the result of which is
shown in Fig. 2(b). The transition is relatively sharp with a
full width at the half maximum of the imaginary-part signal being
$\sim$ 0.8\,K, which is comparable to that found in the
bulk.\cite{Canfield01}

The critical current densities of the MgB$_2$ films were
determined using the standard Bean model\cite{Gyorgy89} from dc
magnetization curves measured with a Quantum Design PPMS
magnetometer. In Fig. 3, the temperature dependence of $J_c$ is
plotted for a 4000\,\AA-thick MgB$_2$ film. A zero-field $J_c
\sim 1.34 \times 10^6$\,A/cm$^2$ was obtained at 7.5\,K. The $M-H$
loop at 10\,K for magnetic field $\mathbf{H} \perp$ film surface,
which is 5 mm $\times$ 4 mm in size, is shown in the inset. It
shows a severe instability in flux pinning, or flux jump, which
causes the collapse of circulating critical current so that the
magnetization curve returns to the reversible magnetization
branch (in the II and IV quadrants). The pinning instability is
absent when $\mathbf{H} \parallel$ film surface, but the small
film thickness yields erroneously large $J_c$ values from the
critical-state model. The actual closing of the hysteresis curve
excluding the flux jump occurs at a field of $\sim$ 8\,T at low
temperatures, suggesting an irreversibility field similar to that
found in bulk MgB$_2$.\cite{Larbalestier01}

The $T_{c0}$ value reported here is much higher than those in the
early reports of {\em in situ} MgB$_2$ thin
films.\cite{Blank01,Christen01,Shinde01,Grassano01} This
demonstrates that {\em in situ} annealed MgB$_2$ films can have
$T_{c0}$ values close to those of the bulk. The critical current
density ($\sim 10^6$ A/cm$^2$) is also close to those of bulk
MgB$_2$.\cite{Finnemore01,Larbalestier01} These bulk-like
superconducting properties suggest that although the electron
diffraction patterns in Fig. 1 are dominated by the MgO features,
the MgB$_2$ phase is formed during the {\em in situ} annealing
process. The MgB$_2$ grains are believed to be very small (<5 nm),
so that their diffraction spots are not seen in region I (Fig.
1(b)) and weak in region II (Fig. 1(c)).

Compared to the {\em ex situ} annealed films, which are
textured\cite{Shinde01,HYZhai01,Eom01} and have a grain size of
order 10 nm\cite{Eom01}, $J_c$ in our films is an order of
magnitude lower.\cite{WNKang01,Eom01,Moon01} This, as well as the
slightly lower $T_c$ and the severe pinning instability may be
attributed to the small grain size in the {\em in situ} annealed
films. When the grain size is close to the coherence length of
MgB$_2$, the superconducting properties may be affected. While
the 900 $^{\circ}$C annealing in the {\em ex situ} process
provides enough thermal energy for crystallization and texturing,
the lower temperature during the {\em in situ} annealing limits
the extent of these processes. The film deposited at 250 -
300\,$^{\circ}$C was likely a mixture of Mg and amorphous MgB$_2$
or B, and the MgB$_2$ crystallites were formed during the heating
and annealing steps. In this annealing step, the film thickness
changed from $\sim$\,8000\,\AA\, to $\sim$\,4000\,\AA\,
indicating the evaporation of excess Mg and possible
decomposition of MgB$_2$. The annealing temperature and time are
constrained by the thermodynamics\cite{ZKLiu01} and kinetic
barriers\cite{ZYFan01} of the Mg-B system. The fabrication of
high-quality {\em in situ} annealed films requires a delicate
balance between the MgB$_2$ phase formation and decomposition.

In conclusion, we have deposited MgB$_2$ thin films by pulsed
laser deposition using an {\em in situ} annealing process. The
$T_{c0}$ obtained is much higher than the previously reported
{\em in situ} films and $J_c$ is comparable to those of the
polycrystalline bulk samples even though the grain size in the
films is extremely small. Because this deposition process is more
compatible with multilayer deposition, it is important to
demonstrate that high $T_{c0}$ and $J_c$ can be obtained using
this process. Our results show that it can be a viable candidate
for MgB$_2$ Josephson junction technologies.

This work is supported in part by ONR under grant No.
N00014-00-1-0294 (XXX), by NSF under grant Nos. DMR-9875405 and
DMR-9871177 (XQP), DMR-9876266 and DMR-9972973 (QL), DMR-9983532
(ZKL), and by DOE through grant DE-FG02-97ER45638 (DGS). The work
at BNL (QL) is supported by DOE, Office of BES, under contracts
No. DE-AC02-98CH10886.


\begin{figure}[h]
 \epsfig{figure=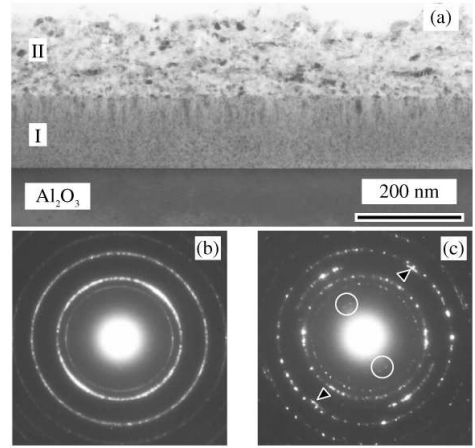, width = 8cm}
 \caption{ (a) Dark-field TEM image showing a
cross-sectional view of a MgB$_2$ thin films. Selected-area
electron diffraction patterns from (b) region I and (c) region
II.}
\end{figure}

\begin{figure}[h]
\vspace{1cm}
 \epsfig{figure=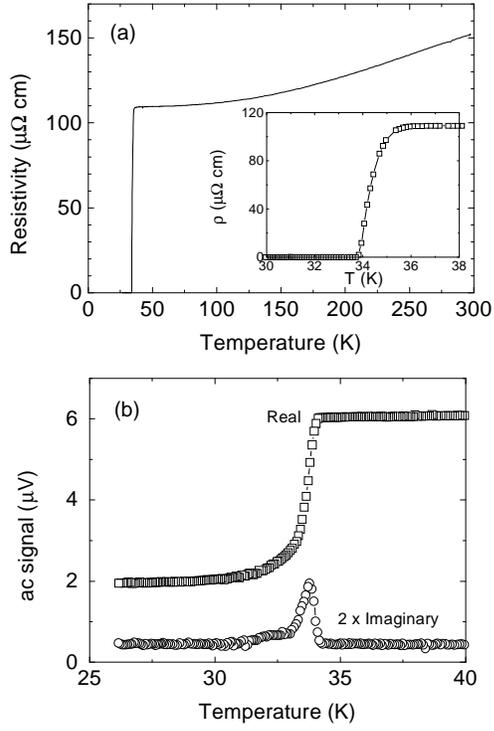, width = 6.5cm }
 \caption{ (a) Resistivity versus temperature curve for a 4000 \AA-thick MgB$_2$
film. (b) The ac susceptibility of the same film.}
\end{figure}

\begin{figure}[h]
 \epsfig{figure=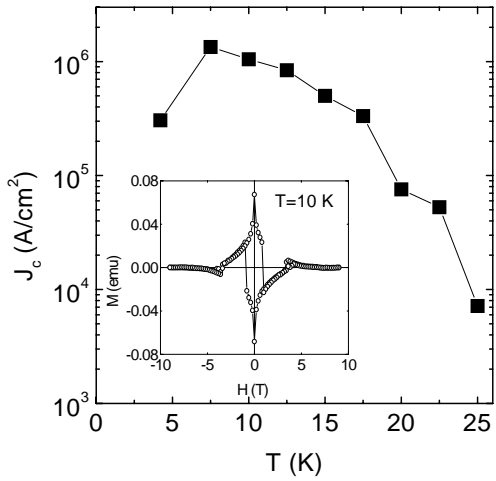, width = 6.5cm } \caption{
The temperature dependence of the zero-field $J_c$ of a 4000
\AA-thick MgB$_2$ film. The inset shows the $M-H$ loop at $T=10$
K.  }
\end{figure}


\begin{thebibliography}{19}
\expandafter\ifx\csname
natexlab\endcsname\relax\def\natexlab#1{#1}\fi
\expandafter\ifx\csname bibnamefont\endcsname\relax
  \def\bibnamefont#1{#1}\fi
\expandafter\ifx\csname bibfnamefont\endcsname\relax
  \def\bibfnamefont#1{#1}\fi
\expandafter\ifx\csname url\endcsname\relax
  \def\url#1{\texttt{#1}}\fi
\expandafter\ifx\csname
urlprefix\endcsname\relax\def\urlprefix{URL }\fi
\providecommand*{\bibinfo}[2]{#2} \providecommand*{\eprint}[1]{#1}
\providecommand*{\url}[1]{#1}
\begingroup\makeatletter
 \@temptokena{%
  \expandafter\ifx\csname citenamefont\endcsname\relax
   \DeclareRobustCommand\citenamefont{\@firstofone}%
   \global\let\citenamefont\citenamefont
   \global\expandafter\let\csname citenamefont \expandafter\endcsname\csname
  citenamefont \endcsname
  \fi
 }\if@filesw\immediate\write\@auxout{\the\@temptokena}\fi
\expandafter\endgroup\the\@temptokena

\bibitem[{\citenamefont{Nagamatsu} \emph{et~al.}(2001)\citenamefont{Nagamatsu,
  Nakagawa, Muranaka, Zenitani, and Akimitsu}}]{Nagamatsu01}
\bibinfo{author}{\bibfnamefont{J.}~\bibnamefont{Nagamatsu}},
  \bibinfo{author}{\bibfnamefont{N.}~\bibnamefont{Nakagawa}},
  \bibinfo{author}{\bibfnamefont{T.}~\bibnamefont{Muranaka}},
  \bibinfo{author}{\bibfnamefont{Y.}~\bibnamefont{Zenitani}}, \bibnamefont{and}
  \bibinfo{author}{\bibfnamefont{J.}~\bibnamefont{Akimitsu}},
  \bibinfo{journal}{Nature} \textbf{\bibinfo{volume}{410}}, \bibinfo{pages}{63}
  (\bibinfo{year}{2001}).

\bibitem[{\citenamefont{Cava}(2001)}]{Cava01}
\bibinfo{author}{\bibfnamefont{R.~J.} \bibnamefont{Cava}},
  \bibinfo{journal}{Nature} \textbf{\bibinfo{volume}{410}}, \bibinfo{pages}{23}
  (\bibinfo{year}{2001}).

\bibitem[{\citenamefont{Bud{$^{\prime}$}ko}
  \emph{et~al.}(2001)\citenamefont{Bud{$^{\prime}$}ko, Lapertot, Petrovic,
  Cunningham, Anderson, and Canfield}}]{Budko01}
\bibinfo{author}{\bibfnamefont{S.~L.} \bibnamefont{Bud{$^{\prime}$}ko}},
  \bibinfo{author}{\bibfnamefont{G.}~\bibnamefont{Lapertot}},
  \bibinfo{author}{\bibfnamefont{C.}~\bibnamefont{Petrovic}},
  \bibinfo{author}{\bibfnamefont{C.~E.} \bibnamefont{Cunningham}},
  \bibinfo{author}{\bibfnamefont{N.}~\bibnamefont{Anderson}}, \bibnamefont{and}
  \bibinfo{author}{\bibfnamefont{P.~C.} \bibnamefont{Canfield}},
  \bibinfo{journal}{Phys. Rev. Lett.} \textbf{\bibinfo{volume}{86}},
  \bibinfo{pages}{1877} (\bibinfo{year}{2001}).

\bibitem[{\citenamefont{Karapetrov}
  \emph{et~al.}(2001)\citenamefont{Karapetrov, Iavarone, K.Kwok, Crabtree, and
  Hinks}}]{Karapetrov01}
\bibinfo{author}{\bibfnamefont{G.}~\bibnamefont{Karapetrov}},
  \bibinfo{author}{\bibfnamefont{M.}~\bibnamefont{Iavarone}},
  \bibinfo{author}{\bibfnamefont{W.}~\bibnamefont{K.Kwok}},
  \bibinfo{author}{\bibfnamefont{G.~W.} \bibnamefont{Crabtree}},
  \bibnamefont{and} \bibinfo{author}{\bibfnamefont{D.~G.} \bibnamefont{Hinks}},
  \bibinfo{journal}{cond-mat/0102312}  (\bibinfo{year}{2001}).

\bibitem[{\citenamefont{Finnemore} \emph{et~al.}(2001)\citenamefont{Finnemore,
  Ostenson, Bud{$^{\prime}$}ko, Lapertot, and Canfield}}]{Finnemore01}
\bibinfo{author}{\bibfnamefont{D.~K.} \bibnamefont{Finnemore}},
  \bibinfo{author}{\bibfnamefont{J.~E.} \bibnamefont{Ostenson}},
  \bibinfo{author}{\bibfnamefont{S.~L.} \bibnamefont{Bud{$^{\prime}$}ko}},
  \bibinfo{author}{\bibfnamefont{G.}~\bibnamefont{Lapertot}}, \bibnamefont{and}
  \bibinfo{author}{\bibfnamefont{P.~C.} \bibnamefont{Canfield}},
  \bibinfo{journal}{Phys. Rev. Lett.} \textbf{\bibinfo{volume}{86}},
  \bibinfo{pages}{2420} (\bibinfo{year}{2001}).

\bibitem[{\citenamefont{Larbalestier}
  \emph{et~al.}(2001)\citenamefont{Larbalestier, Cooley, Rikel, Polyanskii,
  Jiang, Patnaik, Cai, Feldmann, Gurevich, Squitieri, Naus, Eom}
  \emph{et~al.}}]{Larbalestier01}
\bibinfo{author}{\bibfnamefont{D.~C.} \bibnamefont{Larbalestier}},
  \bibinfo{author}{\bibfnamefont{L.~D.} \bibnamefont{Cooley}},
  \bibinfo{author}{\bibfnamefont{M.~O.} \bibnamefont{Rikel}},
  \bibinfo{author}{\bibfnamefont{A.~A.} \bibnamefont{Polyanskii}},
  \bibinfo{author}{\bibfnamefont{J.}~\bibnamefont{Jiang}},
  \bibinfo{author}{\bibfnamefont{S.}~\bibnamefont{Patnaik}},
  \bibinfo{author}{\bibfnamefont{X.~Y.} \bibnamefont{Cai}},
  \bibinfo{author}{\bibfnamefont{D.~M.} \bibnamefont{Feldmann}},
  \bibinfo{author}{\bibfnamefont{A.}~\bibnamefont{Gurevich}},
  \bibinfo{author}{\bibfnamefont{A.~A.} \bibnamefont{Squitieri}},
  \bibinfo{author}{\bibfnamefont{M.~T.} \bibnamefont{Naus}},
  \bibinfo{author}{\bibfnamefont{C.~B.} \bibnamefont{Eom}}, \emph{et~al.},
  \bibinfo{journal}{Nature} \textbf{\bibinfo{volume}{410}},
  \bibinfo{pages}{186} (\bibinfo{year}{2001}).

\bibitem[{\citenamefont{Bugoslavsky}
  \emph{et~al.}(2001)\citenamefont{Bugoslavsky, Perkins, Qi, Cohen, and
  Caplin}}]{Bugoslavsky01}
\bibinfo{author}{\bibfnamefont{Y.}~\bibnamefont{Bugoslavsky}},
  \bibinfo{author}{\bibfnamefont{G.~K.} \bibnamefont{Perkins}},
  \bibinfo{author}{\bibfnamefont{X.}~\bibnamefont{Qi}},
  \bibinfo{author}{\bibfnamefont{L.~F.} \bibnamefont{Cohen}}, \bibnamefont{and}
  \bibinfo{author}{\bibfnamefont{A.~D.} \bibnamefont{Caplin}},
  \bibinfo{journal}{Nature} \textbf{\bibinfo{volume}{410}},
  \bibinfo{pages}{563} (\bibinfo{year}{2001}).

\bibitem[{\citenamefont{Shinde} \emph{et~al.}(2001)\citenamefont{Shinde, Ogale,
  Greene, Venkatesan, Canfield, Bud{$\prime$}ko, Lapertot, and
  Petrovic}}]{Shinde01}
\bibinfo{author}{\bibfnamefont{S.~R.} \bibnamefont{Shinde}},
  \bibinfo{author}{\bibfnamefont{S.~B.} \bibnamefont{Ogale}},
  \bibinfo{author}{\bibfnamefont{R.~L.} \bibnamefont{Greene}},
  \bibinfo{author}{\bibfnamefont{T.}~\bibnamefont{Venkatesan}},
  \bibinfo{author}{\bibfnamefont{P.~C.} \bibnamefont{Canfield}},
  \bibinfo{author}{\bibfnamefont{S.}~\bibnamefont{Bud{$\prime$}ko}},
  \bibinfo{author}{\bibfnamefont{G.}~\bibnamefont{Lapertot}}, \bibnamefont{and}
  \bibinfo{author}{\bibfnamefont{C.}~\bibnamefont{Petrovic}},
  \bibinfo{journal}{cond-mat/0103542}  (\bibinfo{year}{2001}).

\bibitem[{\citenamefont{Zhai} \emph{et~al.}(2001)\citenamefont{Zhai, Christen,
  Zhang, Paranthaman, Cantoni, Sales, Fleming, Christen, and
  Lowndes}}]{HYZhai01}
\bibinfo{author}{\bibfnamefont{H.~Y.} \bibnamefont{Zhai}},
  \bibinfo{author}{\bibfnamefont{H.~M.} \bibnamefont{Christen}},
  \bibinfo{author}{\bibfnamefont{L.}~\bibnamefont{Zhang}},
  \bibinfo{author}{\bibfnamefont{M.}~\bibnamefont{Paranthaman}},
  \bibinfo{author}{\bibfnamefont{C.}~\bibnamefont{Cantoni}},
  \bibinfo{author}{\bibfnamefont{B.~C.} \bibnamefont{Sales}},
  \bibinfo{author}{\bibfnamefont{P.~H.} \bibnamefont{Fleming}},
  \bibinfo{author}{\bibfnamefont{D.~K.} \bibnamefont{Christen}},
  \bibnamefont{and} \bibinfo{author}{\bibfnamefont{D.~H.}
  \bibnamefont{Lowndes}}, \bibinfo{journal}{cond-mat}  (\bibinfo{year}{2001}).

\bibitem[{\citenamefont{Kang} \emph{et~al.}(2001)\citenamefont{Kang, Kim, Choi,
  Jung, and Lee}}]{WNKang01}
\bibinfo{author}{\bibfnamefont{W.~N.} \bibnamefont{Kang}},
  \bibinfo{author}{\bibfnamefont{H.-J.} \bibnamefont{Kim}},
  \bibinfo{author}{\bibfnamefont{E.-M.} \bibnamefont{Choi}},
  \bibinfo{author}{\bibfnamefont{C.~U.} \bibnamefont{Jung}}, \bibnamefont{and}
  \bibinfo{author}{\bibfnamefont{S.-I.} \bibnamefont{Lee}},
  \bibinfo{journal}{cond-mat/0104266}  (\bibinfo{year}{2001}).

\bibitem[{\citenamefont{Eom} \emph{et~al.}(2001)\citenamefont{Eom, Lee, Choi,
  Belenky, Song, Cooley, Naus, Patnaik, Jiang, Rikel, Polyanskii, Gurevich}
  \emph{et~al.}}]{Eom01}
\bibinfo{author}{\bibfnamefont{C.~B.} \bibnamefont{Eom}},
  \bibinfo{author}{\bibfnamefont{M.~K.} \bibnamefont{Lee}},
  \bibinfo{author}{\bibfnamefont{J.~H.} \bibnamefont{Choi}},
  \bibinfo{author}{\bibfnamefont{L.}~\bibnamefont{Belenky}},
  \bibinfo{author}{\bibfnamefont{X.}~\bibnamefont{Song}},
  \bibinfo{author}{\bibfnamefont{L.~D.} \bibnamefont{Cooley}},
  \bibinfo{author}{\bibfnamefont{M.~T.} \bibnamefont{Naus}},
  \bibinfo{author}{\bibfnamefont{S.}~\bibnamefont{Patnaik}},
  \bibinfo{author}{\bibfnamefont{J.}~\bibnamefont{Jiang}},
  \bibinfo{author}{\bibfnamefont{M.~O.} \bibnamefont{Rikel}},
  \bibinfo{author}{\bibfnamefont{A.~A.} \bibnamefont{Polyanskii}},
  \bibinfo{author}{\bibfnamefont{A.}~\bibnamefont{Gurevich}}, \emph{et~al.},
  \bibinfo{journal}{cond-mat/0103425}  (\bibinfo{year}{2001}).

\bibitem[{\citenamefont{Moon} \emph{et~al.}(2001)\citenamefont{Moon, Yun, Lee,
  Kye, Kim, Chung, and Oh}}]{Moon01}
\bibinfo{author}{\bibfnamefont{S.~H.} \bibnamefont{Moon}},
  \bibinfo{author}{\bibfnamefont{J.~H.} \bibnamefont{Yun}},
  \bibinfo{author}{\bibfnamefont{H.~N.} \bibnamefont{Lee}},
  \bibinfo{author}{\bibfnamefont{J.~I.} \bibnamefont{Kye}},
  \bibinfo{author}{\bibfnamefont{H.~G.} \bibnamefont{Kim}},
  \bibinfo{author}{\bibfnamefont{W.}~\bibnamefont{Chung}}, \bibnamefont{and}
  \bibinfo{author}{\bibfnamefont{B.}~\bibnamefont{Oh}},
  \bibinfo{journal}{cond-mat/0104230}  (\bibinfo{year}{2001}).

\bibitem[{\citenamefont{Blank} \emph{et~al.}(2001)\citenamefont{Blank,
  Hilgenkamp, Brinkman, Mijatovic, Rijnders, and Rogalla}}]{Blank01}
\bibinfo{author}{\bibfnamefont{D.~H.~A.} \bibnamefont{Blank}},
  \bibinfo{author}{\bibfnamefont{H.}~\bibnamefont{Hilgenkamp}},
  \bibinfo{author}{\bibfnamefont{A.}~\bibnamefont{Brinkman}},
  \bibinfo{author}{\bibfnamefont{D.}~\bibnamefont{Mijatovic}},
  \bibinfo{author}{\bibfnamefont{G.}~\bibnamefont{Rijnders}}, \bibnamefont{and}
  \bibinfo{author}{\bibfnamefont{H.}~\bibnamefont{Rogalla}},
  \bibinfo{journal}{cond-mat/0103543}  (\bibinfo{year}{2001}).

\bibitem[{\citenamefont{Christen} \emph{et~al.}(2001)\citenamefont{Christen,
  Zhai, Cantoni, Paranthaman, Sales, Rouleau, Norton, Christen, and
  Lowndes}}]{Christen01}
\bibinfo{author}{\bibfnamefont{H.}~\bibnamefont{Christen}},
  \bibinfo{author}{\bibfnamefont{H.}~\bibnamefont{Zhai}},
  \bibinfo{author}{\bibfnamefont{C.}~\bibnamefont{Cantoni}},
  \bibinfo{author}{\bibfnamefont{M.}~\bibnamefont{Paranthaman}},
  \bibinfo{author}{\bibfnamefont{B.}~\bibnamefont{Sales}},
  \bibinfo{author}{\bibfnamefont{C.}~\bibnamefont{Rouleau}},
  \bibinfo{author}{\bibfnamefont{D.}~\bibnamefont{Norton}},
  \bibinfo{author}{\bibfnamefont{D.}~\bibnamefont{Christen}}, \bibnamefont{and}
  \bibinfo{author}{\bibfnamefont{D.}~\bibnamefont{Lowndes}},
  \bibinfo{journal}{cond-mat/0103478}  (\bibinfo{year}{2001}).

\bibitem[{\citenamefont{Grassano} \emph{et~al.}(2001)\citenamefont{Grassano,
  Ramadan, Ferrando, Bellingeri, Marr{\'{e}}, Ferdeghini, Grasso, Putti, Siri,
  Manfrinetti, Palenzona, and Chincarini}}]{Grassano01}
\bibinfo{author}{\bibfnamefont{G.}~\bibnamefont{Grassano}},
  \bibinfo{author}{\bibfnamefont{W.}~\bibnamefont{Ramadan}},
  \bibinfo{author}{\bibfnamefont{V.}~\bibnamefont{Ferrando}},
  \bibinfo{author}{\bibfnamefont{E.}~\bibnamefont{Bellingeri}},
  \bibinfo{author}{\bibfnamefont{D.}~\bibnamefont{Marr{\'{e}}}},
  \bibinfo{author}{\bibfnamefont{C.}~\bibnamefont{Ferdeghini}},
  \bibinfo{author}{\bibfnamefont{G.}~\bibnamefont{Grasso}},
  \bibinfo{author}{\bibfnamefont{M.}~\bibnamefont{Putti}},
  \bibinfo{author}{\bibfnamefont{A.~S.} \bibnamefont{Siri}},
  \bibinfo{author}{\bibfnamefont{P.}~\bibnamefont{Manfrinetti}},
  \bibinfo{author}{\bibfnamefont{A.}~\bibnamefont{Palenzona}},
  \bibnamefont{and}
  \bibinfo{author}{\bibfnamefont{A.}~\bibnamefont{Chincarini}},
  \bibinfo{journal}{cond-mat/0103572}  (\bibinfo{year}{2001}).

\bibitem[{\citenamefont{Canfield} \emph{et~al.}(2001)\citenamefont{Canfield,
  Finnemore, Bud{$^{\prime}$}ko, Ostenson, Lapertot, Cunningham, and
  Petrovic}}]{Canfield01}
\bibinfo{author}{\bibfnamefont{P.~C.} \bibnamefont{Canfield}},
  \bibinfo{author}{\bibfnamefont{D.~K.} \bibnamefont{Finnemore}},
  \bibinfo{author}{\bibfnamefont{S.~L.} \bibnamefont{Bud{$^{\prime}$}ko}},
  \bibinfo{author}{\bibfnamefont{J.~E.} \bibnamefont{Ostenson}},
  \bibinfo{author}{\bibfnamefont{G.}~\bibnamefont{Lapertot}},
  \bibinfo{author}{\bibfnamefont{C.~E.} \bibnamefont{Cunningham}},
  \bibnamefont{and} \bibinfo{author}{\bibfnamefont{C.}~\bibnamefont{Petrovic}},
  \bibinfo{journal}{Phys. Rev. Lett.} \textbf{\bibinfo{volume}{86}},
  \bibinfo{pages}{2423} (\bibinfo{year}{2001}).

\bibitem[{\citenamefont{Gyorgy} \emph{et~al.}(1989)\citenamefont{Gyorgy, {van
  Dover}, Jackson, Schneeneyer, and Waszczak}}]{Gyorgy89}
\bibinfo{author}{\bibfnamefont{E.~M.} \bibnamefont{Gyorgy}},
  \bibinfo{author}{\bibfnamefont{R.~B.} \bibnamefont{{van Dover}}},
  \bibinfo{author}{\bibfnamefont{K.~A.} \bibnamefont{Jackson}},
  \bibinfo{author}{\bibfnamefont{L.~F.} \bibnamefont{Schneeneyer}},
  \bibnamefont{and} \bibinfo{author}{\bibfnamefont{J.~V.}
  \bibnamefont{Waszczak}}, \bibinfo{journal}{Appl. Phys. Lett.}
  \textbf{\bibinfo{volume}{55}}, \bibinfo{pages}{283} (\bibinfo{year}{1989}).

\bibitem[{\citenamefont{Liu} \emph{et~al.}(2001)\citenamefont{Liu, Schlom, Li,
  and Xi}}]{ZKLiu01}
\bibinfo{author}{\bibfnamefont{Z.~K.} \bibnamefont{Liu}},
  \bibinfo{author}{\bibfnamefont{D.~G.} \bibnamefont{Schlom}},
  \bibinfo{author}{\bibfnamefont{Q.}~\bibnamefont{Li}}, \bibnamefont{and}
  \bibinfo{author}{\bibfnamefont{X.~X.} \bibnamefont{Xi}},
  \bibinfo{journal}{Appl. Phys. Lett.}  (\bibinfo{year}{2001}).

\bibitem[{\citenamefont{Fan} \emph{et~al.}(2001)\citenamefont{Fan, Hinks,
  Newman, and Rowell}}]{ZYFan01}
\bibinfo{author}{\bibfnamefont{Z.~Y.} \bibnamefont{Fan}},
  \bibinfo{author}{\bibfnamefont{D.~G.} \bibnamefont{Hinks}},
  \bibinfo{author}{\bibfnamefont{N.}~\bibnamefont{Newman}}, \bibnamefont{and}
  \bibinfo{author}{\bibfnamefont{J.~M.} \bibnamefont{Rowell}},
  \bibinfo{journal}{cond-mat/0103435}  (\bibinfo{year}{2001}).

\end{thebibliography}
\end {document}